\documentclass[pra,twocolumn,notitlepage,superscriptaddress,longbibliography]{revtex4-1}

\usepackage{amsmath}
\usepackage{amssymb}
\usepackage[utf8]{inputenc}
\usepackage[T1]{fontenc}

\usepackage{bm}
\usepackage{color}
\usepackage{epsfig}
\usepackage{physics}
\usepackage{stmaryrd}
\usepackage{textcomp} 

\usepackage[normalem]{ulem}

\graphicspath{{images/}}

\let\(\relax 
\let\)\relax

\newcommand{\eps}{\varepsilon}

\newcommand{\(}{\left(} 
\newcommand{\)}{\right)} 

\usepackage{natbib}
\begin{document}
	\title{Transverse magneto-optical Kerr effect enhanced at the bound states in continuum}
	\author{{V.A. Zakharov$^{1,2}$\email{zakharov.vl.cld@gmail.com} and A.N. Poddubny$^{}$ }}
	\affiliation{$^{}$Ioffe Institute, St. Petersburg 194021, Russia\\
		$^{2}$Skolkovo Institute of Science and Technology, Moscow 121205, Russia}
	\begin{abstract}
		We calculate theoretically the transverse magneto-optical Kerr effect (TMOKE) in the periodically patterned waveguide made from a diluted magnetic semiconductor. It is demonstrated that the TMOKE is resonantly enhanced when the incident wave is in resonance with the bound state in continuum, arising when the quality factor of the leaky waveguide mode increases due to the far-field interference. Our results uncover the potential of all-dielectric and semiconductor nanostructures for resonant magneto-optics.
	\end{abstract}
	\maketitle 
	\section{Introduction}
	Magneto-optical effects, such as the Faraday and the Kerr effect, play a decisive role in the control of light propagation \cite{Zvezdin1997}. In particular, the transverse magneto-optical Kerr effect (TMOKE) can be used to tune the reflected light intensity~\cite{Kerr1878}. The effect is observed for the transverse magnetic (TM)-polarized light, that is incident on a medium magnetized in the direction transverse to the light incidence plane, see Fig.~\ref{fig:1}. However, in traditional bulk metals  magneto-optical effects are weak, and the reflectivity modulation is on the order of $10^{-2}$\textdiv$10^{-4}$ ~\cite{Kerr1878,Krinchik1959,Krinchik1968}. A promising alternative is to exploit the plasmonic resonances of the structure~\cite{Safarov1994,Hermann2001,Belotelov2011,Kreilkamp2013,Murzina2018,Dyakov2018,Spitzer2018,Kolodny2019,Dyakov2019}. For instance, the TMOKE strength up to 1.5\% has been demonstrated in the hybrid nanostructures where the magnetic material is placed near a metallic layer, supporting the surface plasmon resonance \cite{Belotelov2011}. However, the inherent drawback of the plasmonic resonances is the presence of significant Ohmic losses. A possible solution is  to replace metallic structures by   all-dielectric resonant metasurfaces~\cite{Kuznetsov2016,Jahani2016}  that possess only the radiative losses~ \cite{Maksymov2014,Barsukova2019}.
	
	Here, we propose to resonantly enhance the transverse magneto-optical Kerr effect at the so-called optical bound states in continuum (BIC), that have recently attracted a lot of attention \cite{Hsu2013,Hsu2016,Sadrieva2017,Rybin2017,Doeleman2018}. These states arise when the quality factor of the leaky eigenmodes of the periodically patterned dielectric waveguides is boosted due to the destructive interference of the radiation channels, and the leakage is strongly suppressed. Application of the magnetic field shifts the narrow BIC resonance and thus leads to strong relative modulation of the reflection spectrum. 
	
	The paper is organized as follows. In Sec.~\ref{sec:model} we  outline the model and the calculation technique. Numerical results are presented in Sec.~\ref{sec:results}. In Sec.~\ref{sec:analytics} we develop an approximate analytical model, that describes (i) formation of the BIC state  and its manifestation in the reflection spectrum and (ii) its modification by the magnetic field. The model enables us to interpret the numerical results for the dependence of the linewidth and spectral position of the reflection resonance on the incidence angle and applied magnetic field. Our results are summarized in Sec.~\ref{sec:summary}.
	
	\section{Model and numerical approach}\label{sec:model}
	
\begin{figure}[b]
		\includegraphics[width=0.5\columnwidth]{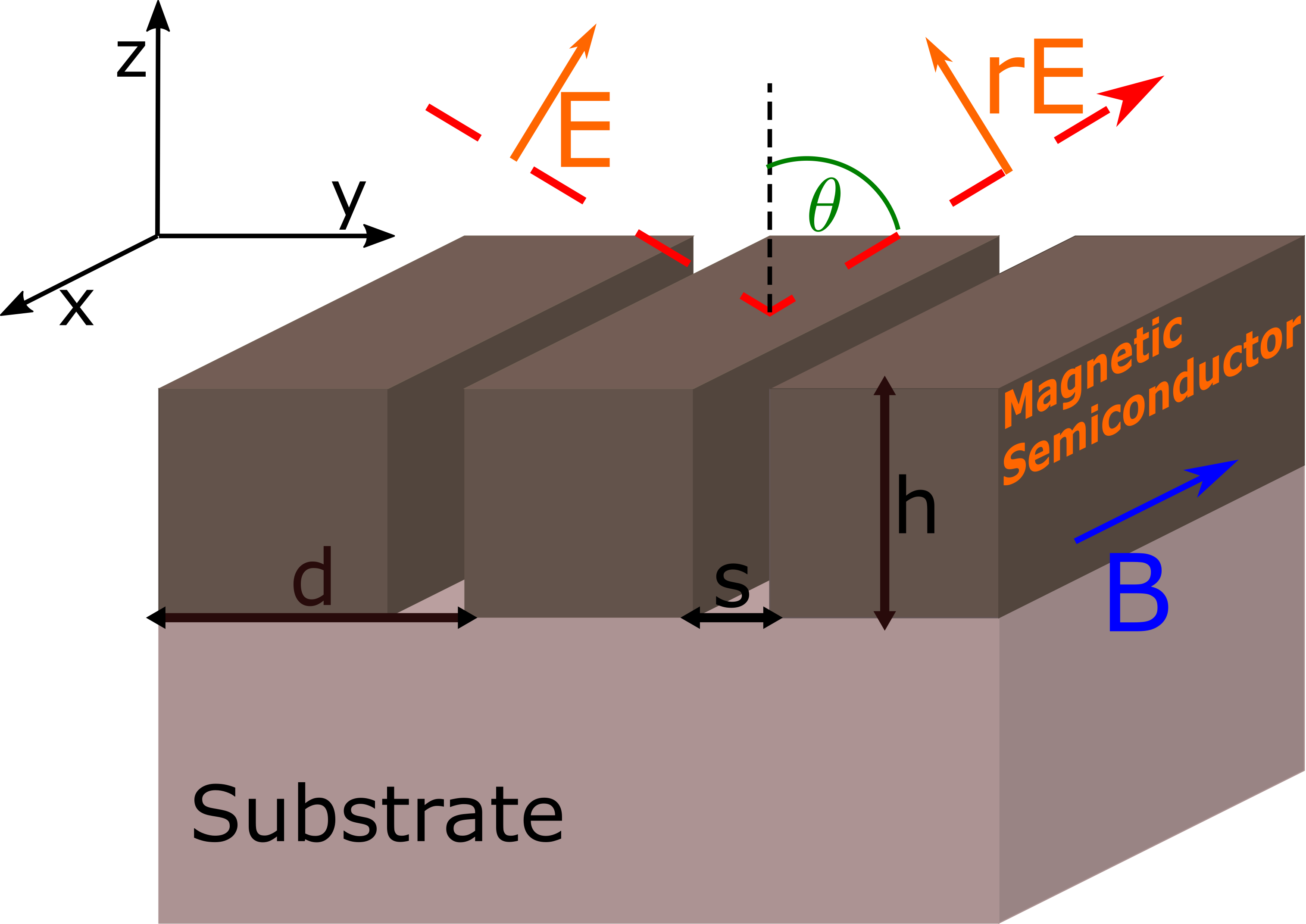}
		\caption{Illustration of the setup to observe the transverse magneto-optical Kerr effect (TMOKE). The structure consists of a semiconductor grating on top of the non-magnetic substrate with  external magnetic field being parallel to the surface and perpendicular to the incidence plane. The incident light is TM-polarized and the incident plane is perpendicular to the grating. The point $z = 0$ corresponds to the middle of the magnetic layer.}\label{fig:1}
	\end{figure}

	The structure under consideration is schematically shown in Fig.~\ref{fig:1}. It consists of a periodic grating made of magnetic semiconductor deposited on the substrate. The grating layer has thickness $h$, grating period is $d$ and the width of the slits in the grating is $w$. The light is obliquely incident under the angle $\theta$ in transverse magnetic (TM) polarization and the incident plane is perpendicular to the grating.  
	External magnetic field is applied along the surface of the sample, perpendicular to the incidence plane.
	The essence of the TMOKE effect is the dependence of the absolute value of the light reflection coefficient on the incidence angle, $R(\theta, B)\equiv |r(\theta,B)|^{2}\ne R(-\theta,B)= R(\theta,-B)$, where $r$ is the ratio of the amplitudes of reflected and incident electromagnetic waves. The TMOKE manifests itself in transmission as well as in reflection, and our results are applicable to both geometries.
	
	In order to obtain the reflection coefficient we expand the electromagnetic field in different layers of the structure over the plane waves,
	solve Maxwell equations for each layer, and then apply the boundary conditions at every boundary.  Li factorization technique is used to improve convergence~\cite{Li1996, Li1998, Popov2001}. 
	This  computational scheme is  extensively used   for a various multilayered structures due to its high numerical efficiency and relative simplicity~\cite{Whittaker1999,Gippius2005,Liscidini2008,Maksymov2014,Ivchenko2014}.
	It is also termed in the literature as a scattering matrix method~\cite{Gippius2005,Liscidini2008} and can viewed as a specific variant of a coupled-modes theory~\cite{Yang2014}.
	The details of the approach are briefly described below. We start by obtaining a basis set of eigenfunctions of the electro-magnetic field propagating inside the magnetic layer, $-h/2<z<h/2$:
	Namely, we look for the solutions to the Maxwell equations for the magnetic field $\vb H$ and electric field $\vb E$ in case of TM polarization in the form
	\begin{align}
	\vb H &= \vb {e_x} e^{ik^{med}_zz}\sum \limits_{n=-\infty}^{\infty}H_ne^{i(\frac{2\pi n}{d}+k_y)y}, \label{eq:H}\\
	\vb E &= e^{ik^{med}_zz}\sum \limits_{n=-\infty}^{\infty}(\vb {e_y} E_{y,n}+\vb {e_z} E_{z,n})e^{i(\frac{2\pi n}{d}+k_y)y}\label{eq:E}.
	\end{align}
	The permittivity tensor in the magnetic grating reads
	\begin{align}
	\hat \eps (y,-h/2<z<h/2)= \begin{pmatrix}
	\eps & 0 &0\\
	0&\eps&igB\\
	0&-igB&\eps
	\end{pmatrix},
	\end{align}
	where $\eps=1$ and $g=0$ between the magnetic rods, when $md<y<s+md, m \in \mathbb{Z}$. Next, we substitute the electromagnetic field \eqref{eq:H}, \eqref{eq:E} into the Maxwell equations 	$\curl{\vb E} = \frac{i\omega}{c}\vb H$, $\curl{\vb H} = - \frac{i\omega}{c}\hat \eps \vb E$.  Taking into account the proper convolution rules for the  Fourier series \cite{Li1998}, the Maxwell equations assume the form
	\begin{gather}\label{eq:maglayer}
	(k_y+\tfrac{2\pi n}{d})E_{z,n}-k^{med}_zE_{y,n} =  \tfrac{\omega}{c} H_n,\\\nonumber
	k^{med}_zH_n=-\frac{\omega}{c}(\llbracket\tfrac{1}{\eps_{22}}\rrbracket^{-1})_{nl}\( E_{y,l}+\llbracket\tfrac{\eps_{23}}{\eps_{22}}\rrbracket_{lm} E_{z,m} \),\nonumber\\
	(k_y+\tfrac{2\pi n}{d})H_n=k^{med}_z\llbracket\tfrac{\eps_{23}}{\eps_{22}}\rrbracket_{nl} H_l+\tfrac{\omega}{c}\llbracket\eps_{22}+\tfrac{\eps_{23}^2}{\eps_{22}}\rrbracket_{nl} E_{z,l}\:.\nonumber
	\end{gather}
	Here, the notation
	\begin{equation}
	\llbracket \eps\rrbracket_{lm}\equiv \frac1{d}\int\limits_{0}^{d} {\rm d} y f(y)e^{i\frac{2\pi}{d}(m-l)y}
	\end{equation}
	is used for the matrix of the Fourier components of the given combination of the components of the dielectric tensor $\hat \eps (y)$.
	The key ingredient of the approach, ensuring the fast numerical convergence, is the judicious order of the Fourier integration and matrix inversion. While  the  series with the factor $(\llbracket\tfrac{1}{\eps_{22}}\rrbracket^{-1})_{nl}$  replaced by	 $\llbracket \eps_{22}\rrbracket_{nl}$ is ill-behaved  at the discontinuities of the permittivity, the form in Eq.~\eqref{eq:maglayer} ensures much faster convergence.
	The truncated system of equations Eq.~\eqref{eq:maglayer} is  solved numerically to determine the wave vector components $k^{med}_{m,z}{\equiv k^{med}_{m}}$ and 
	the corresponding eigenvector components $H^{(m)}_n$. Here, the index $m$ labels the eigensolutions.
	
	At the next stage we write the eigenfunction expansion for the magnetic field in the whole structure. It worth to notice that eigenfunctions for the air and the substrate layers are ordinary plane waves. Magnetic field is parallel to the $x$ axis in $TM$ polarization.
	\begin{align}\label{eq:Hgen}
	H_x =\begin{cases}
	H_0 e^{ik_yy}[e^{-ik^{air}_0(z-\frac{h}{2})} -\!\sum \limits_{n=-\infty}^{\infty}\!r_ne^{i(z-\frac{h}{2})k^{air}_n}e^{i\frac{2\pi n}{d}y}], \\\hfill (z>h/2)\\
	H_0 e^{ik_yy}\sum \limits_{m=1}^\infty [b_mu_m(y) e^{-ik^{med}_{m}z}+c_ms_m(y)e^{ik^{med}_{m}z}], \\ \hfill (-h/2<z<h/2)\\
	H_0e^{ik_yy} \sum \limits_{n=-\infty}^{\infty}t_ne^{-i(z+\frac{h}{2})k^{sub}_n}e^{iy\frac{2\pi n}{d}}, \hfill (z<-h/2)\:.
	\end{cases}
	\end{align}
	Here, we have introduced the $z$-components of the wave vectors in air and substrate  $k^{air}_n=\sqrt{\frac{\omega^2}{c^2}-(k_y+\frac{2\pi n}{d})^2}$, 	$k^{sub}_n = \sqrt{\eps_{sub}\frac{\omega^2}{c^2}-(k_y+\frac{2\pi n}{d})^2}$,  respectively, and $\omega$ is the incident light frequency.
	The coefficients $r_{0}\equiv r$ and $t_{0}\equiv t$ characterize the specular reflection and direct transmission, respectively, while the coefficients with $n\ne 0$ describe the waves diffracted on the grating. The functions $u_{m}(y)$ and $s_{m}(y)$ are given by ${\sum_{n=-\infty}^{\infty}u_m^n e^{i\frac{2\pi}{d}n}}$, ${\sum_{n=-\infty}^{\infty}s_m^n e^{i\frac{2\pi}{d}n}}$ {, respectively, where $u_m^n\equiv H^{(m)}_n, s_m^n\equiv H^{(-m)}_n$ have been obtained from the system 
		Eq.~\eqref{eq:maglayer} for opposite $k^{med}_z$ vector components}. Electric field expansion  over the plane waves is obtained by applying the Maxwell equation $\curl{\vb H} = - \frac{i\omega}{c}\hat \eps \vb E$ to Eq.~\eqref{eq:Hgen}.
	Next, we use the Maxwell boundary conditions of	the continuity of tangential components of electric and magnetic field
	\begin{align}
	(H^{out}_x)_n\big|_{z=\pm \frac{h}{2}} = (H^{med}_x)_n\big|_{z=\pm \frac{h}{2}}\:,\\
	(E^{out}_y)_n\big|_{z=\pm \frac{h}{2}} = (E^{med}_y)_n\big|_{z=\pm \frac{h}{2}}\:.
	\end{align}
	for all the Fourier components. The resulting linear system of equations is solved to find the electromagnetic field in the whole structure.
	\section{Results of calculation}\label{sec:results}
	
	In our modeling we consider a grating made of diluted magnetic semiconductor Cd$_{x}$Mn$_{1-x}$Te on a dielectric BaF$_{2}$ substrate \cite{Janik1996}. We consider the frequency range $\hbar\omega\lesssim 1.6~$eV that is in transparency range. Since the frequencies are relatively close to the band gap edge at $\approx 1.7~$eV, the Verdet constant, characterizing the magneto-optical response, is already relatively large, {$ V~\sim~0.5~\text{deg}/({\text{cm}\cdot\text{Gs}})$ for $T~=~77~\text{K}$} \cite{Gaj1978}.
	The permittivity of Cd$_{x}$Mn$_{1-x}$Te
	$\varepsilon\sim 9$ \cite{Andre1997} is significantly larger than that of the BaF$_{2}$ substrate, $\varepsilon\sim 2$  \cite{Li1980}, hence, the magnetic layer acts as a waveguide.  
	
\begin{figure}[t]
		\includegraphics[width=0.99\columnwidth]{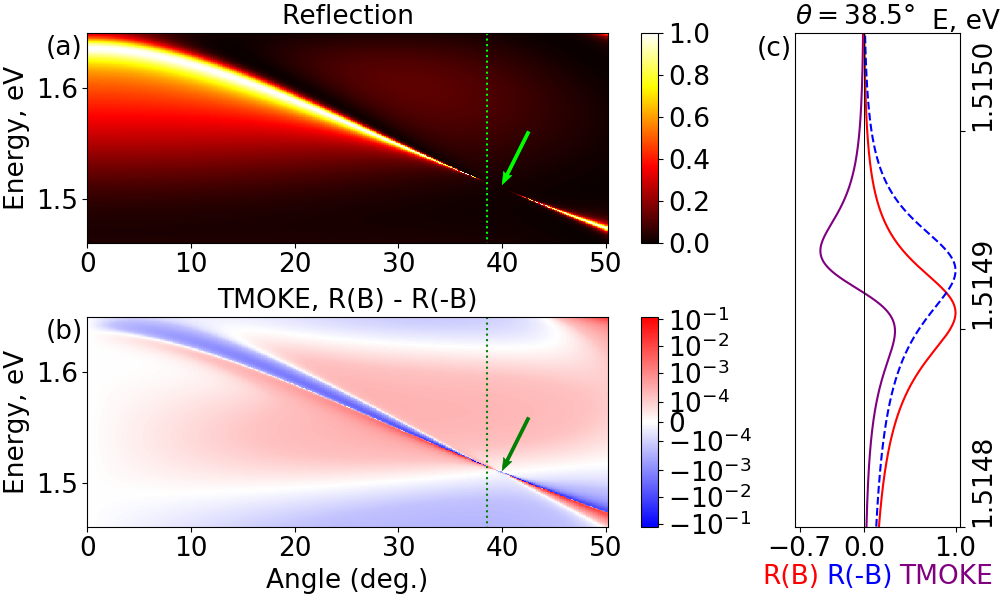}
		\caption{The maps of the reflection coefficient (a) and TMOKE (b) depending on the angle of incidence and on the photon energy. Green arrows point to the leaky waveguide mode that has largest radiative  lifetime (quasi-BIC). Panel (c)  shows two reflection spectra (solid red and dashed blue curve) at a given angle of incidence $\theta=38.5^{\circ}$ 		[indicated by dashed green lines in panels (a) and (b)]
			for opposite magnetic field directions and their difference, $R(B)-R(-B)$, characterizing the  TMOKE strength (dash-dotted purple curve). 
			The parameters of calculation are $ d = 280$~nm, $s = 80$~nm, $h = 560$~nm, $gB = 0.05. $}\label{fig:2}
	\end{figure}

	The calculated light reflection spectra and the TMOKE coefficient are presented in Fig.~\ref{fig:2}.
	Figure~\ref{fig:2}~(a) shows the reflection coefficient dependence on the light energy and incidence angle. Reflectivity demonstrates a distinct maximum, corresponding to the excitation of the leaky waveguide mode in the magnetic layer. In the considered spectral range the structure has no absorption. Thus, the spectral width of the reflection maximum is determined only by the radiative decay. The radiation of the waveguide mode out of the magnetic layer into the far field occurs only through the  zeroth Fourier channel $n=0$, and becomes possible due to the Bragg diffraction on the grating. Importantly, the peak width non-monotonously depends on the incidence angle. The radiative decay is strongly suppressed at the angle $\theta\sim 40^{\circ}$, corresponding to the BIC condition.  This happens because of the destructive interference of
	emission of several leaky eigenmodes of the system (\ref{eq:maglayer}) into the far field \cite{Hsu2013, Hsu2016}. 
	As will be discussed in more detail in the following Section~\ref{sec:analytics}, this interference can be approximately described within the two-mode analytical model. Moreover,   the lifetime remains finite due to the breaking of the mirror symmetry $z\to -z$ by the presence of substrate with $\varepsilon_{\rm subs}\ne 1$ and by the magnetic field, so that the cancellation of the radiative decay is incomplete. However, the width of the reflection peak still reduces  to the values below 1~meV, see Fig.~\ref{fig:2}(c).
	
	The TMOKE coefficient,  $R(B)-R(-B)$, has a resonance and changes sign at the leaky waveguide mode, see Fig.~\ref{fig:2}(b).  Our main result is the strong resonance increase of the TMOKE strength at the BIC condition. The physical origin of this TMOKE enhancement can be understood by explicit comparison of the reflection spectra for opposite values of the magnetic field, shown by red solid and blue dashed curves in Fig.~\ref{fig:2}(c). The calculation demonstrates that there exists a spectral shift of the leaky waveguide mode, that is linear in magnetic field. This results in the corresponding shift of the reflectivity peak, that can be seen by comparing solid red and dashed blue curves. When this peak shift becomes comparable with the peak width, the difference $R(B)-R(-B)$ is strongly increased. Hence, the TMOKE spectrum has a characteristic $S-$like shape, that looks as a derivative from the reflectivity spectrum, see the  purple curve in Fig.~\ref{fig:2}(c). The peak shift weakly depends on the incidence angle, while the peak width is at minimum at the BIC condition, resulting in the sharp maximum of the TMOKE coefficient. 	
	
	\section{Analytical model for magnetic BIC}\label{sec:analytics}
	To describe spectra analytically we developed a two-mode model. In this approximation we keep only zero and minus first components of the Fourier series for all the functions. This is the minimal number of modes necessary to describe the interference responsible for the BIC formation. We needed zeroth Fourier harmonic because it is the only one that propagates outside the waveguide. Choice of the minus first component was justified by the spectral position of the described mode: in the limit of the continuous magnetic layer ($s\to 0$) this mode described only by the minus first component.
	
	First step was to describe the reflectivity and, in particular, the formation of BIC. At this stage there is no need in consideration of the external magnetic field or the substrate. Their effect on the  reflectivity is relatively weak and can be taken into account later as a perturbation. The approach to obtain the reflection and transmission coefficients was the same as described in Sec.~\ref{sec:model}. In the absence of  substrate and  magnetic field the equations are separated into two independent parts: even and odd with respect to the mirror symmetry $z\to -z$. The resulting transmission spectrum consists of the two following  terms:
	\begin{align}\begin{split}
	t_0 & = \(1+\frac{i\Gamma_{0}^{+}(h_+)^{-1}_{-1}}{\det h_+}\) -\(1+ \frac{i\Gamma_{0}^{-}(h_-)^{-1}_{-1}}{\det h_-}\) \\&\equiv\frac{(c_+)_{-1}^{-1}}{\det c_+}-\frac{(c_-)_{-1}^{-1}}{\det c_-}\:.
	\end{split}
	\end{align}
	Here, we use the notation
	\begin{align}
	A_\pm =\begin{pmatrix}
	\Delta^{\pm}_{0}+i\Gamma_0^{\pm}&\kappa^{\pm}\\
	\kappa^{\pm}&\Delta^{\pm}_{-1}
	\end{pmatrix},\\
	c_\pm=A_\pm\left[A_\pm\pm \begin{pmatrix}
	\frac{i}{k^{air}_{0}} & 0\\
	0&\frac{1}{|k^{air}_{-1}|}
	\end{pmatrix}\right]^{-1}, \label{cdef}
	\end{align}
	\begin{align}
	\begin{cases}
	\Delta_{-1}^{+}=\frac{(u^{0}_{-1})^2}{k^{med}_{-1}}\cot\frac{k^{med}_{-1}h}{2}+\frac{(u^{-1}_{0})^2}{k^{med}_{0}}\cot\frac{k^{med}_{0}h}{2}-\frac{1}{\abs{k^{air}_{-1}}},\\
	\Delta^{+}_{0}=\frac{(u^{0}_{0})^2}{k^{med}_{0}}\cot\frac{k^{med}_{0}h}{2}+
	\frac{(u^{-1}_{-1})^2}{k^{med}_{-1}}\cot\frac{k^{med}_{-1}h}{2},\\
	\Gamma^{+}_{0}=\frac{-1}{k^{air}_0},\\
	\kappa^{+}=\frac{u^{-1}_{-1}u^{0}_{-1}}{k^{med}_{-1}}\cot\frac{k^{med}_{-1}h}{2}+\frac{u^{-1}_{0}u^{0}_{0}}{k^{med}_{0}}\cot\frac{k^{med}_{0}h}{2},\label{notation+}
	\end{cases}\end{align}
	\begin{align}
	\begin{cases}
	\Delta_{-1}^{-}=\frac{(u^{0}_{-1})^2}{k^{med}_{-1}}\tan\frac{k^{med}_{-1}h}{2}+\frac{(u^{-1}_{0})^2}{k^{med}_{0}}\tan\frac{k^{med}_{0}h}{2}+\frac{1}{\abs{k^{air}_{-1}}},\\
	\Delta^{-}_{0}=\frac{(u^{0}_{0})^2}{k^{med}_{0}}\tan\frac{k^{med}_{0}h}{2}+
	\frac{(u^{-1}_{-1})^2}{k^{med}_{-1}}\tan\frac{k^{med}_{-1}h}{2},\\
	\Gamma^{-}_{0}=\frac{1}{k^{air}_0},\\
	\kappa^{-}=\frac{u^{-1}_{-1}u^{0}_{-1}}{k^{med}_{-1}}\tan\frac{k^{med}_{-1}h}{2}+\frac{u^{-1}_{0}u^{0}_{0}}{k^{med}_{0}}\tan\frac{k^{med}_{0}h}{2},\label{notation-}
	\end{cases} \end{align}
	where the sign $+$ corresponds to the contribution of the even part of the field, and $-$ to the odd part. We will later omit  the $\pm$ signs when describing the features that hold for both odd and even parts.
	
	By construction of the effective 2-mode ``Hamiltonians'' $A_\pm$, zeroes of the $\Delta_{0,-1}$ function for a given angle of incidence correspond to the energy of the two uncoupled eigenmodes of the waveguide; coefficient $\Gamma_{0}$ respond for the damping of zero mode and the non-diagonal component $\kappa$ describe  the coupling of the two modes due to the Bragg diffraction on a grating.
	
	The condition of observing the BIC is the equality of the $A$ matrix determinant to zero for real frequency, $\omega~\in~\mathbb{R}$, which means that\begin{align}
	\begin{cases}
	\kappa(\omega) = 0, \\
	\Delta_{-1}(\omega) = 0.
	\end{cases}
	\end{align}
	In other words we observed the BIC when the minus first mode is decoupled from the zero mode. Generalization of this rule to the multicomponent calculation with many Fourier harmonics taken into account can be formulated as: ``The BIC is observed when the hybrid leaky waveguide mode, that consists of all the localized Fourier modes, decouples from the zero mode.'' This is a specific case of a more general rule of Friedrich-Wintgen BIC (\hspace{0pt}\mbox{\cite{Hsu2016}},~\mbox{\cite{Friedrich1985}}) for interfering resonances $\sqrt{\Gamma_{0}\Gamma_{-1}}\(\Delta_{-1}-\Delta_{0}\)=\kappa\(\Gamma_{0}-\Gamma_{-1}\)$. In our case we have $\Gamma_{-1} = 0$ because the minus first Fourier channel of radiation can not propagate outside the waveguide and thus has no damping.

	The systems of Eqs.~(\ref{notation+}, \ref{notation-})  can be simplified by employing the linear approximation for $\Delta_{0,-1}(\omega,\theta) = \delta_{0,-1}(\theta)[\omega-\omega_{0,-1}(\theta)]$ and constant approximation for the other parameters $\kappa(\omega, \theta)~=~\kappa(\omega^{\pm}_{-1}, \theta)$, $ \Gamma_{0}(\omega, \theta)=\Gamma_{0}(\omega^{\pm}_{0}, \theta) $. This approximation well  describes reflection in the spectral range close to the resonance. It is then possible to estimate the width of the reflection peak at the half maximum as $\Im{\omega}$, where $\omega$ is the root of the equation $\det A(\omega, \theta) = 0$,
	\begin{align}
	W = \frac{2\kappa^2}{\delta_{0}\delta_{-1}\gamma_{0}} = \frac{2(\kappa'[\omega-\omega_{BIC}])^2}{\delta_{0}\delta_{-1}\gamma_{0}}. \label{width}
	\end{align}
	
	\begin{figure}[t!]
		\includegraphics[width=0.99\columnwidth]{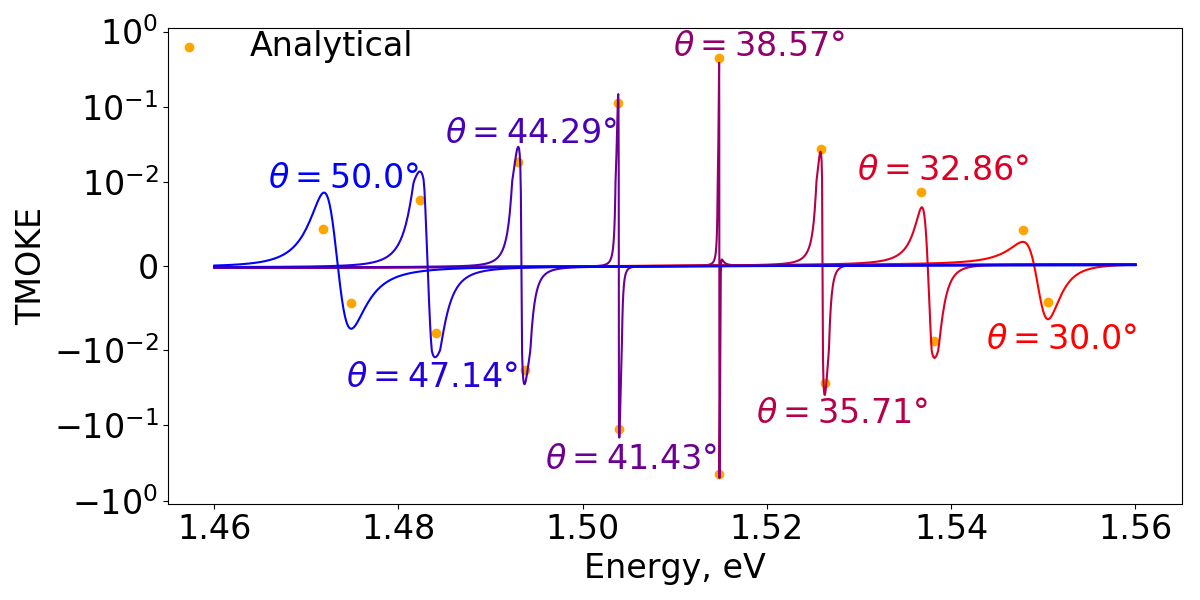}
		\caption{TMOKE coefficient numerically calculated for different angles of incident (curves from red to blue); and analytical approximation of the maximum of the peaks from two-mode model (orange circles).The values of angles are indicated for the corresponding curves.}\label{fig:3}
	\end{figure}
	
	\begin{figure}[b]
		\includegraphics[width=0.99\columnwidth]{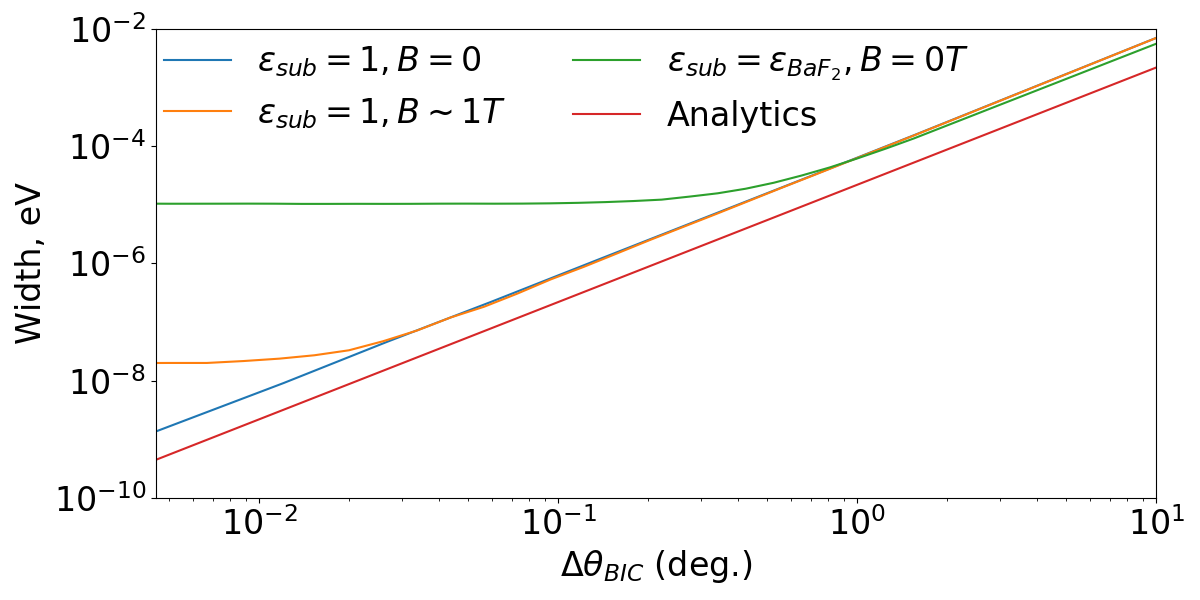}
		\caption{Full width at half maximum  of the reflection peak of the Fabry-Perot resonance  depending on the angle shift from the point with the smallest width. Calculation has been performed numerically for different configurations with/without magnetic field and substrate (solid curves); and  using approximate analytical formula Eq.~\eqref{width} from the  two-mode model (dashed curve).}\label{fig:4}
	\end{figure}

	At the  next step we have taken into account the  effects of external magnetic field and the presence of the substrate on the reflection spectra. It was shown numerically that the linear in magnetic field perturbations in $H^{(m)}_n$ and $k^{med}_z$ are absent, and the only linear effect arises via the perturbation of the electric field, that can be obtained from the Maxwell equation $\curl{\vb H} = - \frac{i\omega}{c}\hat \eps \vb E$. We also do not take into consideration the coupling of different Fourier components of electromagnetic wave by the external magnetic field in the medium. This approximation is reasonable,  because the key effect for TMOKE is just the energy shift of  the waveguide eigenmode.  Hence, we obtain the following formula describing the effect of  the substrate and magnetic field
	\begin{align}
	c_{\pm} \longrightarrow c_{\pm}-M(c_{\mp})^{-1}M\:, \label{corr}
	\end{align}
	where the matrix $M$ is small, $M_{ij} \ll 1$,
	\begin{align}
	M =\begin{pmatrix}
	\frac{k_0^{air}-\frac{k_0^{sub}}{\eps_{sub}}-i\llbracket\frac{gB}{\eps^2}\rrbracket_{00} k_{y}}{2\(k_0^{air}+\frac{k_0^{sub}}{\eps_{sub}}\)}&0\\
	0&\frac{k_{-1}^{air}-\frac{k_{-1}^{sub}}{\eps_{sub}}-i\llbracket\frac{gB}{\eps^2}\rrbracket_{00} \(k_{y}-\frac{2\pi}{d}\)}{2\(k_{-1}^{air}+\frac{k_{-1}^{sub}}{\eps_{sub}}\)}
	\end{pmatrix}. 
	\end{align}
	An important feature of this formula is that it is  quadratic in $M$. If there were no substrate, the matrix $M$ would be proportional to $B$, and hence there would be only quadratic in magnetic field terms in Eq.~\eqref{corr}. In that case TMOKE could not be observed. That means that our model describes an important condition for detecting TMOKE -- the absence of the horizontal mirror symmetry plane of structure \cite{Zvezdin1997}.
	
	From Eq. (\ref{corr}) and Eq. (\ref{cdef}) it is possible to calculate the corrections to $A$ matrix due to the presence of substrate and magnetic field and estimate the frequency shift of the peak as $\Delta \omega = \delta (A_+)_{-1}^{-1}/\delta_{-1}$. Then we can estimate peak value of the TMOKE as the ratio of the frequency shift and the width of the peak $\Delta \omega/W$. Fig.~\ref{fig:3} presents  the comparison of such analytical approximation with the numerical calculation. It demonstrates that our simple analytical model well describes the spectral dependence of TMOKE by the order of the magnitude.  The spectra of TMOKE become very narrow and their amplitude increases near the BIC condition. In case of incidence angle $\theta=38.57^\circ$, the maximum value of TMOKE is equal to 0.5. 
	This value is limited only by the finite detuning from the resonance condition; we expect that the TMOKE amplitude tends to unity when the angle is tuned closer to the resonance.
	
	There is another interesting effect of the external magnetic field and the substrate in addition to the spectral shift of the waveguide eigenmode. They  both break the horizontal mirror symmetry of the system. This induces the coupling between even and odd radiation channels, as  can be seen from (\ref{corr}). As a result, the destructive interference condition necessary to suppress the far-field radiation no longer holds exactly, and the BIC is partially destroyed. Since this effect is weak, the narrow peak in the spectrum survives, but its width does not completely reach zero. These results were proven numerically. Fig.~\ref{fig:4} shows the calculated width of the reflectivity peak for three different configurations in the absence and presence of the magnetic field or substrate, and also the analytical approximation~(\ref{width}). These findings are consistent with the general result that the time-reversal symmetry is necessary for formation of BIC \cite{Hsu2013}.
	
	Figure~\ref{fig:4} demonstrates, that  the minimal linewidth of the BIC state is limited from below by the values $\sim 10~\mu$eV in the presence of BaF$_2$ substrate (green curve) and $\sim 10~$neV in the presence of 1~T external magnetic field. The  corresponding quality factors are  on the order of $10^5$ and $10^8$, and, while being quite high, are in principle relevant for some state-of-the-art experimental structures.
	In fact, the maximal quality factor observed in the seminal experiment \cite{Hsu2013} was about $Q\sim 10^6$ and limited by the spectrometer. Since the $Q$-factor remains finite  due to the substrate or external magnetic field,  it is more rigorous to term the considered resonant state  as a quasi-BIC state, rather than a genuine BIC. However, Fig.~\ref{fig:4} shows that as soon as the angle is detuned from the resonant condition by a small value of $\Delta \theta_{\rm BIC}\sim 1^\circ$, the effects of substrate and external magnetic field become weaker than the radiative losses and the distinction between quasi-BIC and BIC is no longer important.
	Moreover,  the fabrication  quality might strongly affect the lifetime in practice. For example,  the effect of surface roughness has been estimated in \cite{Sadrieva2017} for another kind of  quasi-BIC resonance, that occurs at the normal incidence. The resulting maximum roughness-limited quality factors were predicted to have the values $Q\sim 10^3\divisionsymbol 10^5$ for the effective roughness amplitude changing from $10$~nm to $0.5$~nm, respectively, which is still higher than the value  $Q\sim 300$ observed in experiment of Ref.~\cite{Sadrieva2017}.   As such, we expect that in many realistic cases the quality factor will be  rather be limited by  experimental imperfections, rather than symmetry breaking due to finite magnetic field or a substrate.

	\section{Summary}\label{sec:summary}
	
	To summarize, we studied both numerically and analytically the enhancement of transverse magneto-optical Kerr effect at the resonances of periodically patterned waveguide. We have focused at  the resonance of the bound state in continuum (BIC).  Numerical calculation has been carried out for the realistic material parameters of the diluted magnetic semiconductor Cd$_{x}$Mn$_{1-x}$Te on a dielectric BaF$_{2}$ substrate. 	We demonstrate that the TMOKE strength significantly increases when the  radiative damping of the leaky waveguide mode is quenched due  to the BIC condition. In the narrow frequency range near the BIC resonance the TMOKE strength can reach values up to $1$. The results of constructed two-mode analytical model quantitatively describe the numerical calculation. We have also analyzed the effect of transverse magnetic field and the substrate on the lifetime of the quasibound states in the continuum.
	
	\acknowledgments
	We acknowledge useful discussions with
	I.A. Akimov, D.A. Smirnova and E.L. Ivchenko.
	This work has been financially supported by the Russian Foundation for Basic Research Grant No.~19-52-12038-NNIO\_a. ANP has been partially supported by the Foundation for the Advancement of Theoretical Physics and Mathematics ``BASIS''.
	
	%
\end{document}